\documentclass[twocolumn,tightenlines,prl,floatfix,nofootinbib]{revtex4}

 \usepackage[dvips,final]{graphicx}
  \usepackage{amssymb}
   \usepackage{amsmath}
    \usepackage{amsfonts}
     \usepackage{epsfig}
      \usepackage{bm}

\begin{document}

\title{Finite-width effects in the near-threshold
$ZZZ$ and $ZWW$ production at ILC}

\author{R.~S.~Pasechnik$^1$}
\author{V.~I.~Kuksa$^2$}
 \affiliation{$^1$Uppsala University, Box 516, SE-751 20 Uppsala, Sweden}
 \affiliation{$^2$Institute of Physics, Southern Federal University,
  344090 Rostov-on-Don, Russia}

\begin{abstract}
We calculate the cross-section of the near-threshold off-shell $ZZZ$
and $ZW^+W^-$ production at the International Linear Collider taking
into account their instability and the principal part of NLO
corrections. The calculations are performed in the framework of the
model of unstable particles with smeared mass-shell. We show that
the contribution of the finite $Z/W$ and $H$ widths (their
instability) is large in the Higgs resonance range and should be
taken into account in the Higgs boson searches at future colliders.
\end{abstract}


\maketitle

A great amount of work has been done so far in precision tests of
the Standard model (SM) including measurements of gauge boson, top
quark masses and widths at LEP II \cite{Alcaraz07,Barate03} and
Tevatron \cite{Zhu09}, and very recently at much higher energies at
LHC (see e.g. Ref.~\cite{LHC}). Due to clean environment and
energies well above the electro-weak (EW) scale, future linear
colliders would provide important tools for high-precision
investigation of gauge bosons and Higgs physics in the SM and beyond
\cite{ILC07,ILD10}.

The multiple production of the gauge bosons is crucial for probing
gauge boson (and Higgs) self-couplings, and thus for testing the
non-Abelian structure and EW symmetry breaking of the SM.
The processes of two- ($ZZ$ and $W^+W^-$) and three-boson ($ZZZ$ and
$ZW^+W^-$) production are of major importance as they give a direct
information on trilinear and quartic vector boson couplings.

Triple couplings of the neutral ($Z$ and $\gamma$) and charged
($W^{\pm}$) EW bosons, which were measured at LEP II
\cite{OPAL00,DELPHI07} and Tevatron \cite{D007}, demonstrated a good
agreement with the SM prediction within a few percent
\cite{Alcaraz07}. For this purpose, the NLO EW {\it factorizable}
corrections and {\it finite-width effects} (FWE) in the off-shell
boson pair production i.e. $e^+e^-\to V^*V^* \to 4f$ are very
important, especially, in the near-threshold energy region (see e.g.
Refs.~\cite{Muta86,Denner88}). However, corresponding higher-order
calculations in the framework of traditional perturbation theory
(PT) are rather cumbersome for $2\to4$ processes as require
evaluation of a few thousands one-loop diagrams, and various schemes
for automated loop calculations are practically applied
\cite{Denner07,FC}.

\begin{figure}[h!]
\centerline{\hspace{-4.7cm}\epsfig{file=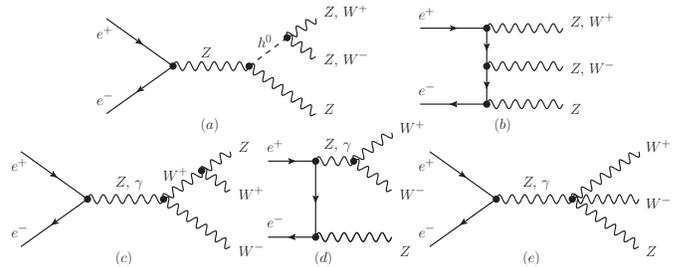,width=4.1cm}}
\caption{Diagrams for $e^{+}e^{-}\to ZZZ, ZW^{+}W^{-}$ processes.}
\label{fig:LO}
\end{figure}

Triple massive gauge boson ($ZZZ$ and $ZW^+W^-$) production
processes can be utilized to probe quartic gauge couplings and
anomalous couplings in the Higgsstrahlung process (see, e.g.
Ref.~\cite{BB,GrosseKnetter92,VVH}). At the moment, these processes
being intensively studied in literature
\cite{Barger88,JiJuan08,Wei09,Boudjema09}. Typical leading-order
contributions are shown in Fig.~\ref{fig:LO}. In this work, we are
primarily concentrated on the three-boson production processes in
the Standard Model at ILC as the simplest case.

Generally, FWE in a multiple gauge bosons production are closely
connected to their instability, so they are usually referred to as
the unstable particles (UP). Near-threshold production of the
unstable particles, as a rule, is accompanied by large FWE, which
must be taken into account in analysis of corresponding observables
\cite{Muta86,Denner88}. In addition to the standard PT approach to
FWE analysis, based on the stable particles approximation (SPA),
where UP instability is accounted for by higher-order corrections,
various approximation schemes are practically applied in the
literature, namely, semi-analytical approximation
\cite{UPtrad,Bardin}, improved Born approximation \cite{IBA},
asymptotical expansions of the cross section in powers of coupling
constant \cite{exp}, fermion loop scheme, etc (see, also
Ref.~\cite{revs,Beenakker:1996kt} and references therein). All above
mentioned methods are based on the traditional quantum field theory
of unstable particles \cite{UPtrad}. At the same time, there are
some alternative approaches for UP description such as the effective
theory of UP \cite{Beneke03}, modified perturbation theory
\cite{modPT} and the smeared-mass unstable particles model
\cite{Salam,SMUP}.

The main feature of the FWE is the ``smearing'' (fuzzing) of the
threshold. In the standard treatment, this effect is described by
taking into account all virtual states of UP, i.e. by its
off-shellness. So, the cross section $\sigma(e^+e^-\to VV)$ is
defined as the cross section of inclusive four-fermion production
$\sigma(e^+e^-\to 4f)$ in the double-pole approximation \cite{DPA},
which selects only diagrams with two nearly resonant $V$ bosons and
the number of contributing graphs is considerably reduced. Such an
approximate description is usually realized with the help of the
dressed UP propagators.

In order to describe FWE in triple boson production we have to
consider full cascade process $e^+e^-\to ZZZ,\,ZW^+W^-\to \sum_f
6f$, where the instability of the off-shell bosons is described by
Breit-Wigner propagators. So far, full NLO calculations were
performed in the case of on-shell bosons only in the stable-particle
approximation, i.e. without taking into account FWE and only for
light SM Higgs boson with masses $M_H=120\,\mbox{GeV}$ and
$150\,\mbox{GeV}$ \cite{JiJuan08,Wei09,Boudjema09}. The
corresponding calculations in the traditional perturbation theory
are rather cumbersome, as requires the complete set of a few
thousands of one-loop diagrams.

In the off-shell case, one encounters a very complicated problem.
Exact NLO $2\to6$ matrix elements would require analytical
evaluation of many tens of thousands loop diagrams, and are not
available at the moment, but they are very important for boson FWE
and Higgs contribution studies in vicinity of the threshold and
Higgs resonance with the biggest cross section.

In this paper, we describe FWEs in the near-threshold triple boson
production $e^+e^-\to Z^*Z^*Z^*,\,Z^*W^{+*}W^{-*}\to \sum_f 6f$
within the framework of the smeared-mass unstable particles (SMUP)
model developed in Ref.~\cite{Salam,SMUP}. In Ref.~\cite{KP09} the
conception of the mass smearing as the main element of the SMUP
model was successfully tested by comparison of its predictions with
LEP II data on the boson-pair production $e^+e^-\to
Z^*Z^*,\,W^{+*}W^{-*}\to \sum_f 4f$ total cross sections and
Monte-Carlo simulations including full next-to-leading order matrix
elements. In the framework of this model, the off-shell vector
bosons $Z,\,W^{\pm}$ are treated as unstable particles, and the
smearing of their mass shell effectively accounts for all-order
propagator type corrections; the principal part of other
factorisable corrections can effectively be taken into account as
suggested in Ref.~\cite{KP09}.

In this work, we present the total (inclusive) cross-sections of the
off-shell $ZZZ,\,ZW^+W^-$ bosons production including FWE and the
principal part of NLO (factorizable) corrections coming from the
initial state radiation (ISR) and fussy-mass-shell unstable vector
bosons in the framework of SMUP model. We found a large finite-width
effect for Higgs masses $M_H\gtrsim 2M_W, 2M_Z$, when the
contribution of Higgs diagrams becomes dominant and very sensitive
to the Higgs boson width. So, this effect should be taken into
consideration and can be applied as an auxiliary tool in Higgs boson
searches at future colliders.

\section{Basics of the SMUP model}

\begin{figure}[h!]
\centerline{\epsfig{file=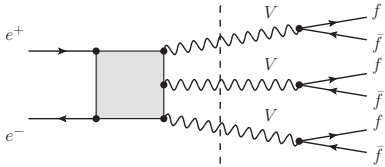,width=4.8cm}} \caption{Schematic
illustration of the exact factorization of the total $ee\to VVV\to
6f$ cross-section in the SMUP model.} \label{fig:fac}
\end{figure}

In the off-shell $ZZZ$ and $ZW^+W^-$ boson production in order to
take into account FWE in standard way we need to consider full
cascade process with six-particle final state and intermediate-state
unstable $Z$ and $W$ bosons (see Fig.~\ref{fig:fac}).

There is another way to tackle the issue. The SMUP model
\cite{Salam,SMUP} provides the possibility to treat $Z$ and $W$
bosons as final state particles and simultaneously to take into
account their instability and obtain the finite-width effects
correctly by smearing their mass shells \cite{KP09,ourev}. In this
Section, we give a short description of the principal elements of
the SMUP model and its advantages compared to the standard treatment
(for more detail review of the SMUP model, see
Ref.~\cite{SMUP,ourev} and references therein).

The UP wave function in the framework of SMUP model is given by
\begin{equation}\label{Phi}
 \Phi_a(x)=\int\Phi_a(x,\mu)\omega(\mu)d\mu,
\end{equation}
where $\Phi_a(x,\mu)$ is the standard spectral component which
defines a particle with a fixed mass squared $m^2=\mu$ in the stable
(fixed-mass) particle approximation (SPA). The weight function
$\omega(\mu)$ is then accounts for the self-energy interactions of
the UP with vacuum fluctuations and decay products. This function
includes all the information about UP decay properties (its
instability) and describes the smeared (``fuzzed'') mass-shell of
the UP. The ``fuzzing'' of the UP mass shell is then caused, on the
one hand, by quantum-mechanical instability according to the
time-energy uncertainty relation and, on the other hand, by
stochastic interactions of the UP with the electro-weak vacuum
fluctuations.

Then, the (anti)commutative relations for the UP field operators
have an additional $\delta$-function in the ``smeared'' UP mass
\cite{SMUP}
\begin{equation}\label{comm}
 [\dot{\Phi}^{-}_{\alpha}({\bar k},\mu),\,\Phi^{+}_{\beta}({\bar q},\mu')]_{\pm}
 =\delta(\mu-\mu') \delta({\bar k}-{\bar q})\delta_{\alpha\beta},
\end{equation}
Here, subscripts ``$\pm$'' correspond to the fermion and boson
fields. The presence of $\delta(\mu-\mu')$ in Eq.~(\ref{comm}) means
that the acts of creation and annihilation of the unstable particles
with different $\mu$ do not interfere. So the quantity $\mu$ has the
status of the physically distinguishable value of the UP mass
squared $m^2$.

In the model under consideration, the transition amplitude of the UP
decay $\Phi\rightarrow\phi_1\phi_2$ directly follows from
Eqs.~(\ref{Phi}) and (\ref{comm}), and can be written as \cite{SMUP}
\begin{equation}\label{Amod}
 A(k,\mu)=\omega(\mu)A^{st}(k,\mu)\,,
\end{equation}
where $A^{st}(k,\mu)$ is the corresponding amplitude in the SPA,
which is calculated in the standard way to a given order of the
Perturbation Theory. From Eq.~(\ref{Amod}) it follows that the
differential (in UP mass squared $\mu$) probability of the
transition is $dP(k,\mu)=\rho(\mu)|A(k,\mu)|^2d\mu$, where
$\rho(\mu)=|\omega(\mu)|^2$ is probability density of mass parameter
$\mu=m^2$. This function is induced by multiple UP interactions with
collective (in our case, EW) vacuum fluctuations of self-energy type
and with decay products (vertex-type corrections). In general, it is
of non-perturbative nature and can be modeled in various ways. In
this work, we use the Lorentz distribution function \cite{SMUP}
\begin{equation}\label{rho}
 \rho(\mu)=\frac{1}{\pi}\,\frac{\mathrm{Im}\,\Pi(\mu)}
 {[\mu-m^2(\mu)]^2+[\mathrm{Im}\,\Pi(\mu)]^2}\,,
\end{equation}
where $\Pi(k^2)$ is the conventional vacuum polarisation function,
$m^2(\mu)=m^2_0+\mathrm{Re}\,\Pi(\mu)$ with bare UP mass $m_0$. The
distribution (\ref{rho}) is of the Breit-Wigner type and accounts
for the electro-weak quantum fluctuations of UP mass shell analogous
to ones leading to a dressing up of the full propagator of an
off-shell particle in the conventional quantum field theory. It
naturally appears in the effective theory of UP as the most suitable
one in the high-energy processes \cite{SMUP}, and was tested before
against LEP II data on off-shell $ZZ$ and $W^+W^-$ near-threshold
production \cite{KP09}.

As a general feature, such a model allows to consider the UP as the
final state particles and their instability is then included by a
convolution of decay widths or cross sections with UP mass
probability density $\rho(m_i^2)$ in each UP leg due to exact
factorisation property which was proven previously in
Ref.~\cite{UPfac}. As a consequence of UP mass-smearing effect, this
factorisation drastically simplifies the calculations. Now, we wish
to apply the formalism of SMUP model for prediction of the
finite-width effects in the triple off-shell ($ZZZ$ and $ZW^+W^-$)
bosons production.

\section{Mass-shell smearing effects
in the triple bosons production}

The processes $e^+e^-\to ZZZ;\,ZW^+W^-$ at the tree level (leading
order) in SPA are described by the set of diagrams represented in
Fig.~\ref{fig:LO}. Here, $ZZZ$ production is described by nine
diagrams with topologies $a,\,b$, whereas $ZW^+W^-$ production is
given by sixteen diagrams $a,...,\,e$. The first subset of diagrams
$a$ has a resonant character at $M_H\gtrsim 2M_W, 2M_Z$, and plays
significant role in the Higgs and FWE contributions. Complete NLO
corrections to on-shell bosons production are described by
additional few thousands diagrams, which were calculated in
Refs.~\cite{JiJuan08,Wei09,Boudjema09}.

Due to exact factorisation property, the SMUP model
\cite{SMUP,UPfac} allows to represent the total cross-section of the
inclusive process $e^+e^-\to ZZZ,\,ZW^+W^-\to \sum_f 6f$ in the
factorized triple-convolution form (as schematically illustrated in
Fig.~\ref{fig:fac})
\begin{eqnarray}\label{ccsec}
&&\sigma(s)=\int\int\int dm^2_1dm^2_2dm^2_3\,
\sigma(s;m^2_1,m^2_2,m^2_3)\\
&&\qquad\qquad\qquad\times\rho(m^2_1)\rho(m^2_2)\rho(m^2_3),
\nonumber
\end{eqnarray}
where $\sigma(s;m^2_1,m^2_2,m^2_3)$ is the Born-level cross-section
as a function of different bosons masses squared $m_i^2,\,i=1,2,3$
in the SPA, $\rho(m_i^2)$ is the probability distribution of the
boson mass squared in $i$th leg given by Eq.~\ref{rho}. The Born
$ZZZ$ and $ZW^+W^-$ production cross-sections were calculated using
FeynCalc v6.1 \cite{FC} as functions of smeared masses of $Z$ and
$W^{\pm}$ bosons $m_i^2$, and then their convolutions with
$\rho(m_i^2)$ over variable $m_i^2$ are performed numerically. In
order to treat the poles in the boson propagators, which arise in
the Higgs resonance region in the integration over the phase space
of the boson pairs, we introduce the $q^2$-dependent decay width of
Higgs boson in the propagator \cite{Bredenstein06}
\begin{equation}\label{Hwidth}
\Gamma_H(q)=\Gamma^{st}_H(q)+\Gamma^{WW}_H(q)+\Gamma^{ZZ}_H(q),
\end{equation}
where $\Gamma^{st}_H(q)$ is standard width of Higgs at the pole
$M^2_H=q^2$ \cite{KP09} and other two terms account for the
boson-pair channels of the Higgs decay. It should be noted here,
that the exact Higgs width as a function of the momentum transfer
scale (\ref{Hwidth}) is very important since it determines to a
large extent the $ZZZ$ and $ZW^+W^-$ production cross sections in
the resonance regions, giving rise to the possibility of probing
Higgs decay properties. If Higgs boson is heavier than 160 GeV and
behaves as predicted by the Standard Model, it decays predominantly
into gauge-boson pairs and subsequently into four light fermions.
And triple boson production at ILC can be rather sensitive to extra
(anomalous) contributions to the $HVV$ coupling from the new physics
\cite{VVH}.

The approach under discussion has a close analogy with the
convolution method \cite{Altarelli00}
and the semi-analytical approximation
\cite{UPtrad,Bardin}. However, the status of these approaches are
different \cite{SMUP,KP09,ourev}. In the framework of the SMUP
model, the expression (\ref{ccsec}) directly follows from the UP
smearing-mass conception, and the function $\rho(m_i^2)$ describes
the probability distribution in UP mass squared \cite{Salam,SMUP}.
Moreover, the definition of the unstable particle field function
(\ref{Phi}) determines the strategy of taking into account the major
part of the higher-order corrections in the near-threshold energy
domain \cite{KP09,ourev}.

\section{Cross sections of $ZZZ$ and $ZW^+W^-$ production at ILC}

In the framework of SMUP model, the FWE were previously studied in
the boson-pair production $e^+e^-\to ZZ,\, W^+W^-,\,ZH,\,Z\gamma$ at
LEP II in Refs.~\cite{KP09,ourev}. In the boson-pair production in
vicinity of the threshold, the NLO EW corrections are dominated by
factorizable corrections to EW couplings, propagator-type
(self-energy) corrections and the soft/hard initial and final state
radiation while boxes, pentagons, etc can become relevant only at
energies far from the threshold ones \cite{Beenakker:1996kt}.

As the next natural step, we apply the same strategy in analysis of
the triple boson production. For this purpose, we perform the
following consistency check -- we compared our $e^+e^-\to ZZZ$ and
$ZW^+W^-$ production cross sections in the SPA including only
initial state radiation and NLO vertex corrections (renormalisation
of the boson couplings) with full NLO results from
Refs.~\cite{Wei09,Boudjema09} in the 50 GeV range nearby to the
threshold and got very close results within a percentage accuracy.
This basically proves that non-factorisable, box and pentagon
diagrams do not significantly contribute to the near-threshold
production cross section, and can be omitted for our purposes. This
is the only approximation we adopt in our calculations.

Then we recalculated the on-shell $ZZZ$ and $ZW^+W^-$ production
cross-sections at fixed $M_H=120\,\,\mbox{GeV}$ in SPA taking into
account Higgs boson FWE. Again, the results coincide with ones
reported in Ref.~\cite{Wei09,Boudjema09}. The contribution of Higgs
diagrams and gauge boson FWE into the total cross-section at
$M_H=120\,\,\mbox{GeV}$ turns out to be relatively small. It is
therefore very instructive to look at these contributions for
heavier Higgs boson.

\begin{figure}[h!]
\centerline{\epsfig{file=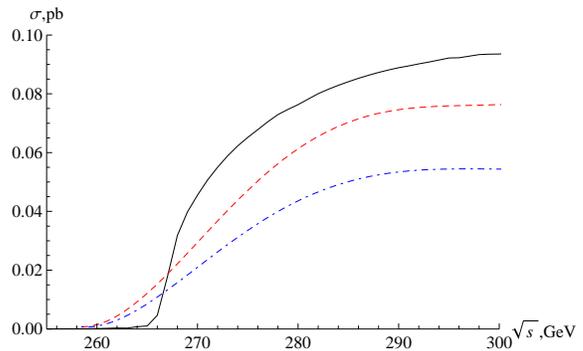,width=7.5cm}}
\caption{Cross-section of the process $e^+e^-\to ZW^+W^-$ at fixed
Higgs mass $M_H=175\,\,\mbox{GeV}$, given at leading order in SPA
(solid line), including gauge bosons FWE (dashed line) and with
taking into account both FWE and ISR corrections (dash-dotted
line).} \label{fig:ZWW-E}
\end{figure}

The cross-section of the process $e^+e^-\to ZW^+W^-$ as function of
$\sqrt{s}$ at fixed $M_H=175\,\,\mbox{GeV}$ (just above the latest
Tevatron exclusion limit \cite{ICHEP}) is shown in
Fig.~\ref{fig:ZWW-E}. The Born cross sections in SPA and with taking
into account the $Z/W$ FWE (due to the instability of gauge bosons
in the final state) in the framework of SMUP model are represented
by solid and dashed lines, respectively. The corresponding
cross-section including the initial state radiation (ISR)
corrections is given by the dashed-dotted line. These corrections
together with the UP propagator-type corrections effectively taken
into account by UP mass-smearing effects in the framework of SMUP
model are the main part of NLO corrections at considered energies
\cite{KP09}\footnote{The major part of the vertex EW corrections at
the threshold can be effectively included by taking coupling
constants at the $M_Z$ mass scale, while contribution from the
renormalisation group evolution is small at energies close to the
threshold \cite{KP09}.}. One can see from Fig.~\ref{fig:ZWW-E} that
the contribution of $Z/W$ FWE is occurred to be quite large for
relatively heavy Higgs boson in the near-threshold region.

\begin{figure}[h!]
\centerline{\epsfig{file=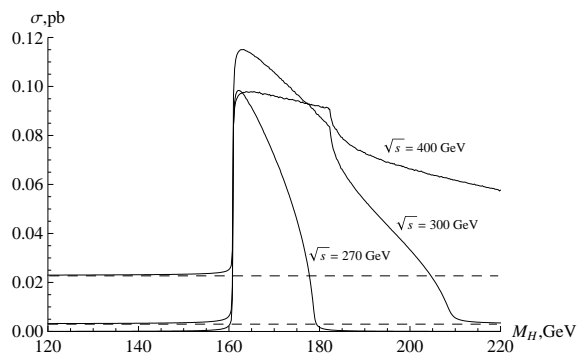,width=7.5cm}}
\caption{Cross-section of the process $e^+e^-\to ZW^+W^-$ as
function of $M_H$ at different energies. Contributions of the
Higgs-less diagrams only are given by dashed lines.}
\label{fig:ZWW-MH}
\end{figure}

In Fig.~\ref{fig:ZWW-MH}, the Born cross-section is shown as
function of Higgs mass $M_H$ at various fixed energies. Solid lines
represent the total cross-section while the dashed lines -- the part
given by Higgs-less diagrams only. This figure illustrates rather
strong dominance of the Higgs contribution due to Higgs resonance in
the $W^+W^-$ phase space. It is worth to note here that the gauge
boson FWE and smearing-mass effects naturally decrease at larger
energies, and get practically negligible at $\sqrt{s}\gtrsim 450$
GeV.

\begin{figure}[h!]
\centerline{\epsfig{file=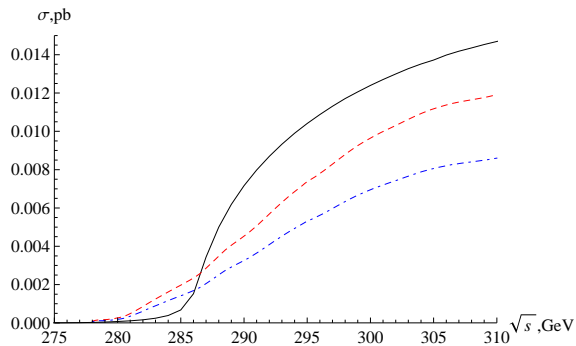,width=7.5cm}}
\caption{Cross-section of the process $e^+e^-\to ZZZ$ at fixed Higgs
mass $M_H=195$ GeV. Notations here are the same as in
Fig.~\ref{fig:ZWW-E}.} \label{fig:ZZZ-E}
\end{figure}

\begin{figure}[h!]
\centerline{\epsfig{file=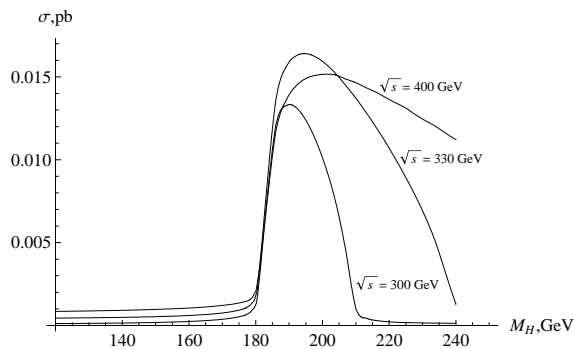,width=7.5cm}}
\caption{Cross-section of the process $e^+e^-\to ZZZ$ as function of
$M_H$ at different energies.} \label{fig:ZZZ-MH}
\end{figure}

Analogously, the cross-section of the process $e^+e^-\to ZZZ$ as
function of $\sqrt{s}$ at fixed $M_H=195\,\,\mbox{GeV}$ close to the
$H\to ZZ$ threshold is given by Fig.~\ref{fig:ZZZ-E}, with the same
notations as before. In Fig.~\ref{fig:ZZZ-MH}, the Born
cross-section is shown as function of Higgs mass at different fixed
energies. Again, similarly to the $ZW^+W^-$ case, the contributions
of gauge bosons and Higgs FWE is large in the near-threshold region
close to the Higgs resonance.

In Conclusion, we would like to notice that using the argument that
the mass-smearing conception realized in the smeared-mass unstable
particles model is in the good agreement with the LEP II
experimental data on the near-threshold boson-pair production, we
applied the same ideas to off-shell $ZZZ$ and $ZW^+W^-$ boson
production at linear collides. The approach under consideration
significantly simplifies the calculations with respect to the
traditional one due to the exact factorization property. Explicit
calculations of the triple-boson production cross sections
demonstrate rather strong dependence of the gauge boson finite-width
effects on the Higgs boson mass. In the near-threshold energy domain
such effects are large and comparable with the initial state
radiation corrections when the Higgs-resonant contribution is
significant, and small when the Higgs contribution is negligible.
The Higgs-resonant contribution into the total cross-section is
strongly dominant and have well-defined signature at Higgs masses
above the Higgs decay threshold $M_H\gtrsim 2M_W$ for $ZW^+W^-$
production and at $M_H\gtrsim 2M_Z$ for $ZZZ$ production.


Useful discussions with Rikard Enberg, Gunnar Ingelman, Oscar
St{\aa}l and Glenn Wouda are gratefully acknowledged. This work was
partially supported by the Carl Trygger Foundation.

\end{document}